\documentclass[12pt]{article}
\usepackage{yfonts}
\usepackage{amsmath}
\usepackage{amssymb}
\usepackage{amstext}
\usepackage{amscd}
\usepackage{amsfonts}
\newcommand{\be}{\begin{equation}}
\newcommand{\nd}{\noindent}
\newcommand{\ee}{\end{equation}}
\newcommand{\ben}{\begin{eqnarray}}
\newcommand{\een}{\end{eqnarray}}

\title{{\bf  Inversion of Tsallis'
 q-Fourier Transform and the complex-plane generalization
}}

\author{ A. Plastino and M. C. Rocca \\
Departamento de F\'{\i}sica-CCT-IFLP- CONICET \\
Fac. de Ciencias Exactas,
Universidad Nacional de La Plata\\
C.C. 67 (1900) La Plata, Argentina}

\date{December 1, 2011}

\begin{document}

\maketitle

\begin{abstract}

We introduce a complex q-Fourier transform as a generalization of
the (real) one analyzed in [Milan J. Math. {\bf 76} (2008) 307].
By recourse to tempered ultradistributions we show that this
complex plane-generalization overcomes all troubles that afflict
its real counterpart.

\end{abstract}

\newpage

\renewcommand{\theequation}{\arabic{section}.\arabic{equation}}

\section{Introduction}

Nonextensive statistical mechanics (NEXT) \cite{[1],[2],AP}, a
current generalization of the Boltzmann�Gibbs (BG) one, is
actively studied in diverse areas of Science. NEXT is based on a
nonadditive (though extensive \cite{[3]}) entropic information
measure characterized by the real index q (with q = 1 recovering
the standard BG entropy). It has been applied to variegated
systems such as cold atoms in dissipative optical lattices
\cite{[4]}, dusty plasmas \cite{[5]}, trapped ions \cite{[6]},
spinglasses \cite{[7]}, turbulence in the heliosheath \cite{[8]},
self-organized criticality \cite{[9]}, high-energy experiments at
LHC/CMS/CERN \cite{[10]} and RHIC/PHENIX/Brookhaven \cite{[11]},
low-dimensional dissipative maps \cite{[12]}, finance \cite{[13]},
galaxies \cite{AP1}, Fokker-Planck equation's applications
\cite{AP2}, etc.

An idiosyncratic NEXT feature is that can it can be advantageously
expressed via q-generalizations of standard mathematical concepts
\cite{borges}. Included are, for instance, the logarithm and
exponential functions, addition and multiplication, Fourier
transform (FT) and the Central Limit Theorem (CLT) \cite{tq2}.
 The q-Fourier transform $F_q$ exhibits the nice property of
 transforming q-Gaussians into q-Gaussians \cite{tq2}.
Recently, plane waves, and the representation of the Dirac delta
into plane waves have been also generalized
\cite{[15],[16],tq1,tq4}.

We will be concerned here with the fact that a generic analytical
expression for the inverse q-FT for arbitrary functions and any
value of q does not exist \cite{tq4}. Investigations revolving
around this fact and related questions might be relevant for field
theory and condensed matter physics, engineering (e.g., image and
signal processing), and mathematical areas for which the standard
FT and its inverse play important roles. It has been recently
shown \cite{FFF} that, in the $1 < q < 2$ particular case, and for
non-negative functions (e.g., probability distributions), it is
possible, by using special kinds of information, to obtain a
bi-univocal relation between a function and its q-FT.

In this work we focus attention on the fact that, for fixed $q$,
the q-Fourier transform is NOT one-to-one. The $F_q-$scenario can
be vastly improved, however, by recourse to tempered
ultradistributions (TUD) \cite{bollini}, that help  generalizing
such transform to the complex plane. The generalization
ameliorates the troubles that (see, for example \cite{FFF})
afflict the real $F_q$.

Why TUD? Because they solve  characterization problems for
analytic functions whose boundary values are elements of the
spaces of distributions, or, conversely, of finding
representations of elements of the quoted spaces of generalized
functions by analytic functions. Many papers concern themselves
with the ultradistribution spaces of Sebastiao e Silva \cite{tp1}.
Such spaces are related to the solvability and the regularity
problems of partial differential equations. Because of such
relation, the study of the structural problems as well as problems
of various operations and integral transformations in this setting
is interesting in itself. Thus, an analysis of spaces of
distributions considered as boundary values of analytic functions
having appropriate growth estimates, is of great value. One wishes
to deal, in particular, with the Dirac's integral representation
in ultradistribution spaces, with the convolution of tempered
ultradistributions and ultradistributions of exponential type (in
Quantum Field Theory), and with the integral transforms of
tempered ultradistributions, of which the best known is the
Fourier complex transformation \cite{bollini}.

\section{The Complex q-Fourier Transform and its Inverse}

So-called q-exponentials \be \label{qexp} e_q(x)=
[1+(1-q)x]_+^{1/(1-q)},\ee are the hallmark of Tsallis's
statistics \cite{[1]}. They are generalizations of the ordinary
exponential functions and coincide with them for $q=1$. We start
our considerations by appealing to a {\it complex} q-exponential.
More precisely, we speak of $e_q(ikx)$ for $1<q<2$ with $k$ a real
number (see Ref. (\cite{tq1})
\begin{equation}
\label{ep1.1} e_q(ikx)=[1+i(1-q)kx]^{\frac {1} {1-q}}.
\end{equation}

It can be seen that  $e_q(ikx)$ is the cut along the real k-axis
of the tempered ultradistribution (see the Appendix for details)
\begin{equation}
\label{ep1.2}
E_q(ikx)=\left\{H(x)H[\Im(k)]-H(-x)H[-\Im(k)]\right\}
[1+i(1-q)kx]^{\frac {1} {1-q}},
\end{equation}
where $H(x)$ is the Heaviside's step function and $\Im(k)$  the
imaginary part of the complex number $k$.

Define now the set $\Lambda_{q,\infty}$,  defined as
\begin{equation}
\label{ep1.4}
{\Lambda}_{q,\infty}=\{f(x)/f(x)\in{\Lambda}_{q,\infty}^+\wedge
f(x)\in{\Lambda}_{q,\infty}^-\},
\end{equation}
where
\[{\Lambda}_{q,\infty}^+=\left\{f(x)/f(x)\{1+i(1-q)kx[f(x)]^{(q-1)}\}^{\frac {1} {1-q}}\in
{\cal L}^1[\mathbb{R}^+]\wedge \right.\]
\begin{equation}
\label{ep1.5} \left. [f(x)\geq 0;1\leq q<2]\right\},
\end{equation}
and
\[{\Lambda}_{q,\infty}^-=\left\{f(x)/f(x)\{1+i(1-q)kx[f(x)]^{(q-1)}\}^{\frac {1} {1-q}}\in
{\cal L}^1[\mathbb{R}^-]\wedge\right.\]
\begin{equation}
\label{ep1.6} \left. [f(x)\geq 0;1\leq q<2]\right\}.
\end{equation}

With the help of this set and using   (\ref{ep1.2}) we define our
complex Tsallis' q-Fourier transform (of $f(x)\in
\Lambda_{q,\infty}$) in the fashion
\[F(k,q)=[H(q-1)-H(q-2)]\times \]
\[\left\{H[\Im(k)]\int\limits_0^{\infty} f(x)\{1+i(1-q)kx[f(x)]^{(q-1)}\}^{\frac {1}
{1-q}},
\;dx -\right.\]
\begin{equation}
\label{ep1.3} \left. H[-\Im(k)]\int\limits_{-\infty}^0 f(x)
\{1+i(1-q)kx[f(x)]^{(q-1)}\}^{\frac {1} {1-q}} \;dx\right\}.
\end{equation}
In (\ref{ep1.3}) $q$ is a real variable such that $1\leq q<2$. It
is of the essence that the cut along the real axis of this
transform is the real Tsallis' q-Fourier transform given in
\cite{tq2}, \cite{tq1}. Taking into account that for $q=1$ the
q-Fourier transform is the usual Fourier transform and using the
formula for the inversion of the complex Fourier transform
immediately yields the inversion formula for (\ref{ep1.3}):
\begin{equation}
\label{ep1.7} f(x)=\frac {1}
{2\pi}\oint\limits_{\Gamma}\left[\lim_{\epsilon\rightarrow 0^+}
\int\limits_1^2 F(k,q)\delta (q-1-\epsilon)\;dq\right]
e^{-ikx}\;dk.
\end{equation}
Eqs.  (\ref{ep1.3}) and (\ref{ep1.7}) solve the problem of
inversion of the q-Fourier transform, {\it which is  now  of the
desired  one-to-one character} (see \cite{tq3},\cite{tq5}) for
fixed $q$. Of course, on the real axis we obtain for (\ref{ep1.3})
and (\ref{ep1.7})
\[F(k,q)=[H(q-1)-H(q-2)]\times \]
\begin{equation}
\label{ep1.8} \int\limits_{-\infty}^{\infty} f(x)
\{1+i(1-q)kx[f(x)]^{(q-1)}\}^{\frac {1} {1-q}} \;dx,
\end{equation}
for the real transform, and
\begin{equation}
\label{ep1.9} f(x)=\frac {1}
{2\pi}\int\limits_{-\infty}^{\infty}\left[\lim_{\epsilon\rightarrow
0^+} \int\limits_1^2 F(k,q)\delta (q-1-\epsilon)\;dq\right]
e^{-ikx}\;dk,
\end{equation}
for its inverse.

\setcounter{equation}{0}

\section{A case in which  $F_q$ is not one-to-one}

Let us discuss an interesting example and  consider Hilhorst's
work \cite{tq4} to illustrate the unfortunate fact that for {\bf
fixed} $q$ the q-Fourier transform is not one-to-one. Let $f(x)$
be given by:
\begin{equation}
\label{ep4.1}
f(x)=
\begin{cases}
\left(\frac {\lambda} {x}\right)^{\beta}\;;\; x\in[a,b]\;;\; 0<a<b\;;\;\lambda>0 \\
0\;;\;x\; \rm{outside}\; [a,b]
\end{cases}
\end{equation}
The corresponding complex q-Fourier transform is:
\[F(k,q)=[H(q-1)-H(q-2)] H[\Im(k)]\times\]
\begin{equation}
\label{ep4.2}
{\lambda}^{\beta}
\int\limits_a^b x^{-\beta}\{1+i(1-q)k{\lambda}^{\beta(q-1)}
x^{1-\beta(q-1)}\}^{\frac {1} {1-q}}
\;dx
\end{equation}
By effecting the change of variable
\[y=x^{1-\beta(q-1)}\]
we obtain for (\ref{ep4.2}):
\[F(k,q)=[H(q-1)-H(q-2)] H[\Im(k)]\times\]
\begin{equation}
\label{ep4.3} \frac {{\lambda}^{\beta}} {1-\beta(q-1)}
\int\limits_{a^{1-\beta(q-1)}}^{b^{1-\beta(q-1)}} y^{\frac
{\beta(q-2)} {1-\beta(q-1)}}\{1+i(1-q)k{\lambda}^{\beta(q-1)}
y\}^{\frac {1} {1-q}} \;dy.
\end{equation}
(\ref{ep4.3}) can be written equivalently as:
\[F(k,q)=[H(q-1)-H(q-2)]H[\Im(k)]\times\]
\[\left\{\left\{H(q-1)-H\left[q-\left(1+\frac {1} {\beta}\right)\right]\right\}\right.\times\]
\[\frac {{\lambda}^{\beta}} {1-\beta(q-1)}
\int\limits_{a^{1-\beta(q-1)}}^{b^{1-\beta(q-1)}}
y^{-\frac {\beta(2-q)} {1-\beta(q-1)}}\{1+i(1-q)k{\lambda}^{\beta(q-1)}
y\}^{\frac {1} {1-q}}
\;dy+\]
\[\left\{H\left[q-\left(1+\frac {1} {\beta}\right)\right]-H(q-2)\right\}\times\]
\begin{equation}
\label{ep4.4} \left.\frac {{\lambda}^{\beta}} {\beta(q-1)-1}
\int\limits_{b^{1-\beta(q-1)}}^{a^{1-\beta(q-1)}} y^{\frac
{\beta(q-2)} {1-\beta(q-1)}}\{1+i(1-q)k{\lambda}^{\beta(q-1)}
y\}^{\frac {1} {1-q}} \;dy\right\}.
\end{equation}
We appeal now the some of the results to be found in Ref.
(\cite{tt3}) to evaluate (\ref{ep4.4})
\[\int\limits_{a^{1-\beta(q-1)}}^{\infty}
y^{-\frac {\beta(2-q)} {1-\beta(q-1)}}\{1+i(1-q)k{\lambda}^{\beta(q-1)}
y\}^{\frac {1} {1-q}}
\;dy=\]
\[\frac {(q-1)[1-\beta(q-1)]a^{\frac {q-2} {q-1}}} {(2-q)
[(1-q)ik{\lambda}^{\beta}]^{\frac {1} {q-1}}}\times \]
\[F\left(\frac {1} {q-1},\frac {2-q} {(q-1)[1-\beta(q-1)]},
\frac {1} {q-1} + \frac {\beta(2-q)} {1-\beta(q-1)};\right.\]
\begin{equation}
\label{ep4.5} \left.-\frac {1}
{(1-q)ik{\lambda}^{\beta(q-1)}a^{1-\beta(q-1)}}\right),
\end{equation}
and
\[\int\limits_0^{a^{1-\beta(q-1)}}
y^{\frac {\beta(2-q)} {\beta(q-1)-1}}\{1+i(1-q)k{\lambda}^{\beta(q-1)}
y\}^{\frac {1} {1-q}}
\;dy=\]
\[\frac {[\beta(q-1)-1]a^{1-\beta}} {\beta-1}\times \]
\[F\left(\frac {1} {q-1},\frac {\beta-1} {\beta(q-1)-1},
\frac {\beta q-2} {\beta(q-1)-1};\right.\]
\begin{equation}
\label{ep4.6}
\left.(q-1)ik{\lambda}^{\beta(q-1)}a^{1-\beta(q-1)}\right),
\end{equation}
where $F(a,b,c.z)$ is the hypergeometric function.  Thus we obtain
for $F(k,q)$:
\[F(k,q)=[H(q-1)-H(q-2)]H[\Im(k)]\times\]
\[\left\{\left\{H(q-1)-H\left[q-\left(1+\frac {1} {\beta}\right)\right]\right\}\right.\times\]
\[\frac {(q-1){\lambda}^{\beta}} {(2-q)
[(1-q)ik{\lambda}^{\beta}]^{\frac {1} {q-1}}}\times \]
\[\left\{a^{\frac {q-2} {q-1}}F\left(\frac {1} {q-1},\frac {2-q} {(q-1)[1-\beta(q-1)]},
\frac {1} {q-1} + \frac {\beta(2-q)} {1-\beta(q-1)};\right.\right.\]
\[\left.\frac {1} {(q-1)ik{\lambda}^{\beta(q-1)}a^{1-\beta(q-1)}}\right)-\]
\[ b^{\frac {q-2} {q-1}}F\left(\frac {1} {q-1},\frac {2-q} {(q-1)[1-\beta(q-1)]},
\frac {1} {q-1} + \frac {\beta(2-q)} {1-\beta(q-1)};\right.\]
\[\left.\left.\frac {1} {(q-1)ik{\lambda}^{\beta(q-1)}b^{1-\beta(q-1)}}\right)\right\}+\]
\[\left\{H\left[q-\left(1+\frac {1} {\beta}\right)\right]-H(q-2)\right\}
\frac {{\lambda}^{\beta}} {\beta-1}\times\]
\[\left\{a^{1-\beta}F\left(\frac {1} {q-1},\frac {\beta-1} {\beta(q-1)-1},
\frac {\beta q-2} {\beta(q-1)-1};\right.\right.\]
\[\left.(q-1)ik{\lambda}^{\beta(q-1)}a^{1-\beta(q-1)}\right)-\]
\[b^{1-\beta}F\left(\frac {1} {q-1},\frac {\beta-1} {\beta(q-1)-1},
\frac {\beta q-2} {\beta(q-1)-1};\right.\]
\begin{equation}
\label{ep4.7}
\left.\left.\left.(q-1)ik{\lambda}^{\beta(q-1)}b^{1-\beta(q-1)}\right)\right\}\right\}.
\end{equation}
From (\ref{ep4.7}) we appreciate a crucial fact:  our $F(k,q)$
does {\it depend} on $a$ and $b$ (remember that we have shown in
section 2 that $F(k,q)$ is one to one). That the mentioned
dependence is of the essence will become evident right now. If we
fix $q$ and select $\beta=1/(q-1)$ we immediately reproduce the
result obtained by Hilhorst. In fact, in this case we obtain for
(\ref{ep4.7}):
\[F(k,q)={\lambda}^{\frac {1} {q-1}}\frac {q-1} {2-q}H[\Im(k)]
 \left[H(q-1)-H(q-2)\right]\times\]
\[\left[a^{\frac {q-2} {q-1}} F\left(\frac {1} {q-1},\frac {2-q} {q-1},
\frac {2-q} {q-1}; (q-1)ik{\lambda}\right)\right.- \]
\begin{equation}
\label{ep4.8}
\left.b^{\frac {q-2} {q-1}} F\left(\frac {1} {q-1},\frac {2-q} {q-1},
\frac {2-q} {q-1};(q-1)ik{\lambda}\right)\right]
\end{equation}
According to ref.\cite{tt4}:
\[F(-a,b,b,-z)=(1+z)^a.\]
Then, (\ref{ep4.8}) simplifies to:
\[F(k,q)={\lambda}^{\frac {1} {q-1}}\frac {q-1} {2-q}H[\Im(k)]\left[H(q-1)-H(q-2)\right].\]
\begin{equation}
\label{ep4.9} \left(a^{\frac {q-2} {q-1}}-b^{\frac {q-2}
{q-1}}\right) \left[1+(1-q)ik\lambda\right]^{\frac {1} {1-q}}.
\end{equation}
If we use the value of $\lambda$ given by Hilhorst, then
\[\lambda=\left[\left(\frac {q-1} {2-q}\right)
\left(a^{\frac {q-2} {q-1}}-b^{\frac {q-2} {q-1}}
\right)\right]^{1-q}.\]   We now obtain
\begin{equation}
\label{ep4.10} F(k,q)=H[\Im(k)]\left[H(q-1)-H(q-2)\right]
\left[1+(1-q)ik\lambda\right]^{\frac {1} {1-q}}.
\end{equation}
We see  according to (\ref{ep4.10}) that in this case $F(k,q)$ is
{\it independent of $a$ and $b$}, i.e., the same for all
legitimate pairs $a$-$b$, and, as consequence,  not one-to-one
{\sf for fixed} $q$. On the real axis (\ref{ep4.10}) takes de form
\[F(k,q)=\left[H(q-1)-H(q-2)\right]
\left[1+(1-q)i(k+i0)\lambda\right]^{\frac {1} {1-q}}=\].
\begin{equation}
\label{ep4.11} \left[H(q-1)-H(q-2)\right]
\left[1+(1-q)ik\lambda\right]^{\frac {1} {1-q}},
\end{equation}
which is the result obtained in Ref. (\cite{tq4}).

\section{Another example}

As a second example we evaluate the q-Fourier transform of the
Heaviside function
\[f(x)=H(x).\]
In this case,
\begin{equation}
\label{ep4.12} F(k,q)=H[\Im(k)]\int\limits_0^\infty
\left[1+(1-q)ikx\right]^{\frac{ 1} {1-q}}\;dx.
\end{equation}
Using the result given in \cite{tt5} we have
\begin{equation}
\label{ep4.13} F(k,q)=H[\Im(k)]\frac {\Gamma\left(\frac {2-q}
{q-1}\right)} {\Gamma\left(\frac {1}
{q-1}\right)}\left[(1-q)ik\right]^{-1},
\end{equation}
and, finally,
\begin{equation}
\label{ep4.14} F(k,q)=\frac {i} {2-q} \frac {H[\Im(k)]} {k}.
\end{equation}
In the same way, if we select:
\[f(x)=H(-x),\]
we obtain the result:
\begin{equation}
\label{ep4.15} F(k,q)=\frac {i} {2-q} \frac {H[-\Im(k)]} {k}.
\end{equation}
Taking into account that $H(x)+H(-x)=1$ we have,  for
\[f(x)=1,\]
the expression
\begin{equation}
\label{ep4.16} F(k,q)=\frac {i} {2-q} \frac {1} {k}= \frac {2\pi}
{2-q} \delta(k),
\end{equation}
which is the formula obtained by us in Ref.  (\cite{tq1}).

\section*{Conclusions}

Using tempered ultradistributions we have introduced a complex
q-Fourier transform $F(k,q)$ which exhibits nice properties and is
one-to-one.

\nd This solves a serious flaw of the original $F_q-$definition,
i.e., not being of the essential one-to-one nature, as illustrated
in detail in Section 3.

\nd In this work we have shown that if we eliminate the
requirement that $q$ be fixed and let it �float� in its proper
interval $[1,2)$, the complex generalization of the
$F_q-$definition restores the one-to-one character.

\newpage

\section{Appendix: Tempered Ultradistributions and
Distributions of Exponential Type }

\setcounter{equation}{0}

For the benefit of the reader we give a brief summary of the main
properties of distributions of exponential type and tempered
ultradistributions.

\nd {\bf Notations}. The notations are almost textually taken from
Ref. \cite{tp2}. Let $\boldsymbol{{\mathbb{R}}^n}$ (res.
$\boldsymbol{{\mathbb{C}}^n}$) be the real (resp. complex)
n-dimensional space whose points are denoted by
$x=(x_1,x_2,...,x_n)$ (resp $z=(z_1,z_2,...,z_n)$). We shall use
the notations:

(a) $x+y=(x_1+y_1,x_2+y_2,...,x_n+y_n)$\; ; \;
    $\alpha x=(\alpha x_1,\alpha x_2,...,\alpha x_n)$

(b)$x\geqq 0$ means $x_1\geqq 0, x_2\geqq 0,...,x_n\geqq 0$

(c)$x\cdot y=\sum\limits_{j=1}^n x_j y_j$

(d)$\mid x\mid =\sum\limits_{j=1}^n \mid x_j\mid$

\nd Let $\boldsymbol{{\mathbb{N}}^n}$ be the set of n-tuples of
natural numbers. If $p\in\boldsymbol{{\mathbb{N}}^n}$, then
$p=(p_1, p_2,...,p_n)$, and $p_j$ is a natural number, $1\leqq
j\leqq n$. $p+q$ stands for $(p_1+q_1, p_2+q_2,..., p_n+q_n)$ and
$p\geqq q$ means $p_1\geqq q_1, p_2\geqq q_2,...,p_n\geqq q_n$.
$x^p$ entails $x_1^{p_1}x_2^{p_2}... x_n^{p_n}$. We shall denote
by $\mid p\mid=\sum\limits_{j=1}^n p_j $ and call $D^p$  the
differential operator
${\partial}^{p_1+p_2+...+p_n}/\partial{x_1}^{p_1}
\partial{x_2}^{p_2}...\partial{x_n}^{p_n}$

\nd For any natural $k$ we define $x^k=x_1^k x_2^k...x_n^k$ and
${\partial}^k/\partial x^k= {\partial}^{nk}/\partial x_1^k\partial
x_2^k...\partial x_n^k$

\nd The space $\boldsymbol{{\cal H}}$  of test functions such that
$e^{p|x|}|D^q\phi(x)|$ is bounded for any $p$ and $q$, being
defined [see Ref. (\cite{tp2})] by means of the countably set of
norms
\begin{equation}
\label{ep2.1}
{\|\hat{\phi}\|}_p=\sup_{0\leq q\leq p,\,x}
e^{p|x|} \left|D^q \hat{\phi} (x)\right|\;\;\;,\;\;\;p=0,1,2,...
\end{equation}

\nd The space of continuous linear functionals defined on
$\boldsymbol{{\cal H}}$ is the space
$\boldsymbol{{\Lambda}_{\infty}}$ of the distributions of the
exponential type given by ( ref.\cite{tp2} ).
\begin{equation}
\label{ep2.2}
T=\frac {{\partial}^k} {\partial x^k}
\left[ e^{k|x|}f(x)\right]
\end{equation}
where $k$ is an integer such that $k\geqq 0$ and $f(x)$ is a
bounded continuous function. \nd In addition we have
$\boldsymbol{{\cal H}}\subset\boldsymbol{{\cal S}}
\subset\boldsymbol{{\cal S}^{'}}\subset
\boldsymbol{{\Lambda}_{\infty}}$, where $\boldsymbol{{\cal S}}$ is
the Schwartz space of rapidly decreasing test functions
(ref\cite{tp6}).

The Fourier transform of a function $\hat{\phi}\in \boldsymbol{{\cal H}}$
is
\begin{equation}
\label{ep3.1}
\phi(z)=\frac {1} {2\pi}
\int\limits_{-\infty}^{\infty}\overline{\hat{\phi}}(x)\;e^{iz\cdot x}\;dx
\end{equation}
According to ref.\cite{tp2}, $\phi(z)$ is entire analytic and
rapidly decreasing on straight lines parallel to the real axis. We
shall call $\boldsymbol{{\cal H}}$ the set of all such functions.
\begin{equation}
\label{ep3.2} \boldsymbol{{\cal H}}={\cal
F}\left\{\boldsymbol{{\cal H}}\right\}
\end{equation}
The topology in $\boldsymbol{{\cal H}}$ is defined by the
countable set of semi-norms:
\begin{equation}
\label{ep3.4} {\|\phi\|}_{k} = \sup_{z\in V_k} |z|^k|\phi (z)|,
\end{equation}
where $V_k=\{z=(z_1,z_2,...,z_n)\in\boldsymbol{{\mathbb{C}}^n}:
\mid Im z_j\mid\leqq k, 1\leqq j \leqq n\}$

\nd The dual of $\boldsymbol{{\cal H}}$ is the space
$\boldsymbol{{\cal U}}$ of tempered ultradistributions [see Ref.
(\cite{tp2} )]. In other words, a tempered ultradistribution is a
continuous linear functional defined on the space
$\boldsymbol{{\cal H}}$ of entire functions rapidly decreasing on
straight lines parallel to the real axis. \nd Moreover, we have
$\boldsymbol{{\cal H}}\subset\boldsymbol{{\cal S}}
\subset\boldsymbol{{\cal S}^{'}}\subset \boldsymbol{{\cal U}}$.

$\boldsymbol{{\cal U}}$ can also be characterized in the following
way [see Ref. (\cite{tp2} )]: let $\boldsymbol{{\cal A}_{\omega}}$
be the space of all functions $F(z)$ such that:

${\Large {\boldsymbol{A)}}}$-
$F(z)$ is analytic for $\{z\in \boldsymbol{{\mathbb{C}}^n} :
|Im(z_1)|>p, |Im(z_2)|>p,...,|Im(z_n)|>p\}$.

${\Large {\boldsymbol{B)}}}$-
$F(z)/z^p$ is bounded continuous  in
$\{z\in \boldsymbol{{\mathbb{C}}^n} :|Im(z_1)|\geqq p,|Im(z_2)|\geqq p,
...,|Im(z_n)|\geqq p\}$,
where $p=0,1,2,...$ depends on $F(z)$.

\nd Let $\boldsymbol{\Pi}$ be the set of all $z$-dependent
pseudo-polynomials, $z\in \boldsymbol{{\mathbb{C}}^n}$. Then
$\boldsymbol{{\cal U}}$ is the quotient space

${\Large {\boldsymbol{C)}}}$-
$\boldsymbol{{\cal U}}=\boldsymbol{{\cal A}_{\omega}/\Pi}$

\nd By a pseudo-polynomial we understand a function of $z$ of the
form $\;\;$ $\sum_s z_j^s G(z_1,...,z_{j-1},z_{j+1},...,z_n)$ with
$G(z_1,...,z_{j-1},z_{j+1},...,z_n)\in\boldsymbol{{\cal
A}_{\omega}}$

\nd Due to these properties it is possible to represent any
ultradistribution as [see Ref. (\cite{tp2} )]
\begin{equation}
\label{ep3.6}
F(\phi)=<F(z), \phi(z)>=\oint\limits_{\Gamma} F(z) \phi(z)\;dz
\end{equation}
$\Gamma={\Gamma}_1\cup{\Gamma}_2\cup ...{\Gamma}_n,$  where the
path ${\Gamma}_j$ runs parallel to the real axis from $-\infty$ to
$\infty$ for $Im(z_j)>\zeta$, $\zeta>p$ and back from $\infty$ to
$-\infty$ for $Im(z_j)<-\zeta$, $-\zeta<-p$. ($\Gamma$ surrounds
all the singularities of $F(z)$).

\nd Eq. (\ref{ep3.6}) will be our fundamental representation for a
tempered ultradistribution. Sometimes use will be made of the
``Dirac formula" for ultradistributions [see Ref. (\cite{tp1})]
\begin{equation}
\label{ep3.7}
F(z)=\frac {1} {(2\pi i)^n}\int\limits_{-\infty}^{\infty}
\frac {f(t)} {(t_1-z_1)(t_2-z_2)...(t_n-z_n)}\;dt
\end{equation}
where the ``density'' $f(t)$ is such that
\begin{equation}
\label{ep3.8} \oint\limits_{\Gamma} F(z) \phi(z)\;dz =
\int\limits_{-\infty}^{\infty} f(t) \phi(t)\;dt.
\end{equation}
While $F(z)$ is analytic on $\Gamma$, the density $f(t)$ is in
general singular, so that the r.h.s. of (\ref{ep3.8}) should be interpreted
in the sense of distribution theory.

\nd Another important property of the analytic representation is
the fact that on $\Gamma$, $F(z)$ is bounded by a power of $z$
\cite{tp2}
\begin{equation}
\label{ep3.9} |F(z)|\leq C|z|^p,
\end{equation}
where $C$ and $p$ depend on $F$.

\nd The representation (\ref{ep3.6}) implies that the addition of
a pseudo-polynomial $P(z)$ to $F(z)$ does not alter the
ultradistribution:
\[\oint\limits_{\Gamma}\{F(z)+P(z)\}\phi(z)\;dz=
\oint\limits_{\Gamma} F(z)\phi(z)\;dz+\oint\limits_{\Gamma}
P(z)\phi(z)\;dz\] However,
\[\oint\limits_{\Gamma} P(z)\phi(z)\;dz=0.\]
As $P(z)\phi(z)$ is entire analytic in some of the variables $z_j$
(and rapidly decreasing), we obtain:
\begin{equation}
\label{ep3.10}
\oint\limits_{\Gamma}
\{F(z)+P(z)\}\phi(z)\;dz= \oint\limits_{\Gamma} F(z)\phi(z)\;dz.
\end{equation}

\vskip 4mm

\nd {\bf Acknowledgments} The authors thank Prof. C. Tsallis for
having called our attention to the present problem.


\newpage

\begin{thebibliography}{99}

\bibitem{[1]} C. Tsallis, J. Stat. Phys. 52 (1988) 479.
\bibitem{[2]} M. Gell-Mann, C. Tsallis (Eds.), Nonextensive
Entropy: Interdisciplinary Applications, Oxford University Press,
New York, 2004; C. Tsallis, Introduction to Nonextensive
Statistical Mechanics: Approaching a Complex World, Springer, New
York, 2009.
\bibitem{AP} A. R. Plastino, A. Plastino, Phys. Lett A {\bf 177}
(1993) 177.
\bibitem{[3]} C. Tsallis, M. Gell-Mann, Y. Sato, Proc. Natl. Acad. Sci. USA
102 (2005) 15377; F. Caruso, C. Tsallis, Phys. Rev. E 78 (2008)
021102.
\bibitem{[4]} P. Douglas, S. Bergamini, F. Renzoni, Phys. Rev. Lett. 96
(2006) 110601; G.B. Bagci, U. Tirnakli, Chaos 19 (2009) 033113.
\bibitem{[5]} B. Liu, J. Goree, Phys. Rev. Lett. 100 (2008) 055003.
\bibitem{[6]} R.G. DeVoe, Phys. Rev. Lett. 102 (2009) 063001.
\bibitem{[7]} R.M. Pickup, R. Cywinski, C. Pappas, B. Farago, P. Fouquet,
Phys. Rev. Lett. 102 (2009) 097202.
\bibitem{[8]} L.F. Burlaga, N.F. Ness, Astrophys. J. 703 (2009) 311.
\bibitem{[9]} F. Caruso, A. Pluchino, V. Latora, S. Vinciguerra, A.
Rapisarda, Phys. Rev. E 75 (2007) 055101(R); B. Bakar, U.
Tirnakli, Phys. Rev. E 79 (2009) 040103(R); A. Celikoglu, U.
Tirnakli, S.M.D. Queiros, Phys. Rev. E 82 (2010) 021124.
\bibitem{[10]} V. Khachatryan, et al., CMS Collaboration, J. High Energy
Phys. 1002 (2010) 041; V. Khachatryan, et al., CMS Collaboration,
Phys. Rev. Lett. 105 (2010) 022002.
\bibitem{[11]} Adare, et al., PHENIX Collaboration, Phys. Rev. D 83 (2011)
052004; M. Shao, L. Yi, Z.B. Tang, H.F. Chen, C. Li, Z.B. Xu, J.
Phys. G 37 (8) (2010) 085104.
\bibitem{[12]} M.L. Lyra, C. Tsallis, Phys. Rev. Lett. 80 (1998) 53; E.P.
Borges, C. Tsallis, G.F.J. Ananos, P.M.C. de Oliveira, Phys. Rev.
Lett. 89 (2002) 254103; G.F.J. Ananos, C. Tsallis, Phys. Rev.
Lett. 93 (2004) 020601; U. Tirnakli, C. Beck, C. Tsallis, Phys.
Rev. E 75 (2007) 040106(R); U. Tirnakli, C. Tsallis, C. Beck,
Phys. Rev. E 79 (2009) 056209.
\bibitem{[13]} L. Borland, Phys. Rev. Lett. 89 (2002) 098701.
\bibitem{AP1} A. R. Plastino, A. Plastino, Phys. Lett A {\bf 174}
(1993) 834.
\bibitem{AP2} A. R. Plastino, A. Plastino, Physica A {\bf 222}
(1995) 347.
\bibitem{borges} E. P. Borges, Physica A {\bf 340} (2004) 95.
\bibitem{tq2} S. Umarov, C. Tsallis, S. Steinberg, Milan J. Math. 76
(2008) 307; S. Umarov, C. Tsallis, M. Gell-Mann, S. Steinberg, J.
Math. Phys. 51 (2010) 033502.
\bibitem{bollini}  hep-th,  arXiv:hep-th/0309271 (2003).
\bibitem{[15]} M. Jauregui, C. Tsallis, J. Math. Phys. 51 (2010) 063304.
\bibitem{[16]} A. Chevreuil, A. Plastino, C. Vignat, J. Math. Phys. 51
(2010) 093502.
\bibitem{[17]} M. Mamode, J. Math. Phys. 51 (2010) 123509.
\bibitem{tq1} A. Plastino and M.C.Rocca: J. Math. Phys {\bf 52},
103503 (2011).
\bibitem{tq4} H.J.Hilhorst: J. Stat. Mech. P10023 (2010)
\bibitem{tq3} M.Jauregui and C.Tsallis: Phys. Lett. A
{\bf 375}, 2085 (2011).
\bibitem{tq5} M.Jauregui, C.Tsallis and E.M.F. Curado:
arXiv:1108.2690v1.
\bibitem{FFF}   M. Jauregui, C, Tsallis, Phys. Lett. A {\bf 375} (2011) 2085.
\bibitem{tt3} L. S. Gradshtein  and I. M. Ryzhik : {\it Table of Integrals, Series, and
Products}. Fourth edition, Academic Press (1965) 3.194 1 and 3.194 2  pages
284 and 285.
\bibitem{tt4} M.Abramowitz and I.A.Stegun: {\it Handbook of Mathematical
Functions}. National Bureau of Standards. Applied Mathematical Series 55
Tenth Printing (1972), 15.1.8 page 556.
\bibitem{tt5} L. S. Gradshtein  and I. M. Ryzhik : {\it Table of Integrals, Series, and
Products}. Fourth edition, Academic Press (1965) 3.194 3  page
285.
\bibitem{tp1} J. Sebastiao e Silva : Math. Ann. {\bf 136},
38 (1958).
\bibitem{tp2} M. Hasumi: T$\rm{\hat{o}}$hoku Math. J. {\bf 13},
94 (1961).
\bibitem{tp6} L. Schwartz : {\it Th\'eorie des distributions}.
Hermann, Paris (1966).

\end{thebibliography}
\end{document}